\documentclass[twocolumn,english]{revtex4-1}
\usepackage[T1]{fontenc}
\usepackage{color}
\usepackage{amsmath}
\usepackage{amssymb}
\usepackage{graphicx}
\usepackage{esint}
\usepackage{lineno}

\definecolor{samblue}{rgb}{0.0,0.2,0.5}
\definecolor{samgreen}{rgb}{0,0.5,0.0}
\definecolor{samred}{rgb}{0.66,0,0}

\usepackage[normalem]{ulem}

\makeatletter

\renewcommand{\fnum@figure}{\textbf{Figure~\thefigure}}
\usepackage{units}
\usepackage{lipsum}
\usepackage{babel}

\renewcommand{\vec}[1]{{\mathbf{#1}}}

\makeatother

\usepackage{babel}
\begin{document}

\title{Terahertz fingerprint of monolayer Wigner crystals}

\author{Samuel Brem$^1$}
\author{Ermin Malic$^{1,2}$}
\affiliation{$^1$Philipps University, Department of Physics, Marburg, Germany}
\affiliation{$^2$Chalmers University of Technology, Department of Physics, Gothenburg, Sweden}

\begin{abstract}
The strong Coulomb interaction in monolayer semiconductors represents a unique opportunity for the realization of Wigner crystals without external magnetic fields. In this work, we predict that the formation of monolayer Wigner crystals can be detected by their terahertz response spectrum, which exhibits a characteristic sequence of internal optical transitions. We apply the density matrix formalism to derive the internal quantum structure and the optical conductivity of the Wigner crystal and to microscopically analyse the multi-peak shape of the obtained terahertz spectrum. Moreover, we predict a characteristic shift of the peak position as function of charge density for different atomically thin materials and show how our results can be generalized to an arbitrary two-dimensional system.  
\end{abstract}
\maketitle
At low temperatures electrons arrange in a crystal lattice \cite{wigner1934interaction} in order to minimize their repulsive energy, which represents one of the most intriguing quantum phase transitions. Ever since their prediction, the experimental realization of these Wigner crystals (WCs) has remained challenging, considering that the prerequisite of a dominating Coulomb versus kinetic energy requires very low temperatures and charge densities. The first realizations of WCs \cite{andrei1988observation,goldman1990evidence,williams1991conduction} have therefore used strong magnetic fields and quasi two-dimensional (2D) quantum well systems in order to quench the kinetic energy via Landau quantization. 
The advances in the research on transition metal dichalcogenide (TMD) monolayers \cite{wang2018colloquium,mueller2018exciton,raja2019dielectric} have delivered new opportunities to realize electron crystallization. Several recent studies\cite{xu2020correlated,wang2020correlated, zhou2021bilayer,huang2021correlated,miao2021strong} have reported insulating states at fractional fillings of flat moire bands\cite{yu2017moire, tran2019evidence, brem2020tunable}, which arise in the super-lattices of twisted homo- and hetero-bilayer systems. However, similar to the application of magnetic fields, these so-called generalized Wigner crystals heavily rely on the external moire potential stabilizing the lattice structure and can therefore not be considered as intrinsic electronic phases. 

In contrast, TMDs in their monolayer form already represent a unique platform to achieve electron crystallization, as they exhibit large effective masses and a strong Coulomb interaction leading to prominent exciton physics\cite{ugeda2014giant,chernikov2014exciton}. Smole\'nski et al have recently demonstrated signatures of a WC formed in hBN-encapsulated MoSe$_2$ monolayers\cite{smolenski2021signatures} at electron densities of up to $n=3\cdot 10^{11} \text{cm}^{-2}$ without any external fields. As an indicator for the crystallization the authors have used the emergence of an additional peak in the reflectance spectrum at visible frequencies. This peak stems from excitons with finite center-of-mass momentum that couple to the light cone via Bragg scattering at the WC. Other  rather indirect confirmations of the WC, such as a drop in the conductivity, have also been applied to verify other WC systems \cite{andrei1988observation,williams1991conduction,xu2020correlated,zhou2021bilayer,huang2021correlated,goldman1990evidence,miao2021strong}.  However, a direct, more reliable and conclusive experimental proof of the WC, such as a direct spectral signature of the internal structure of the crystal, could not be provided so far.

\begin{figure}[t!]
\includegraphics[width=0.9\columnwidth]{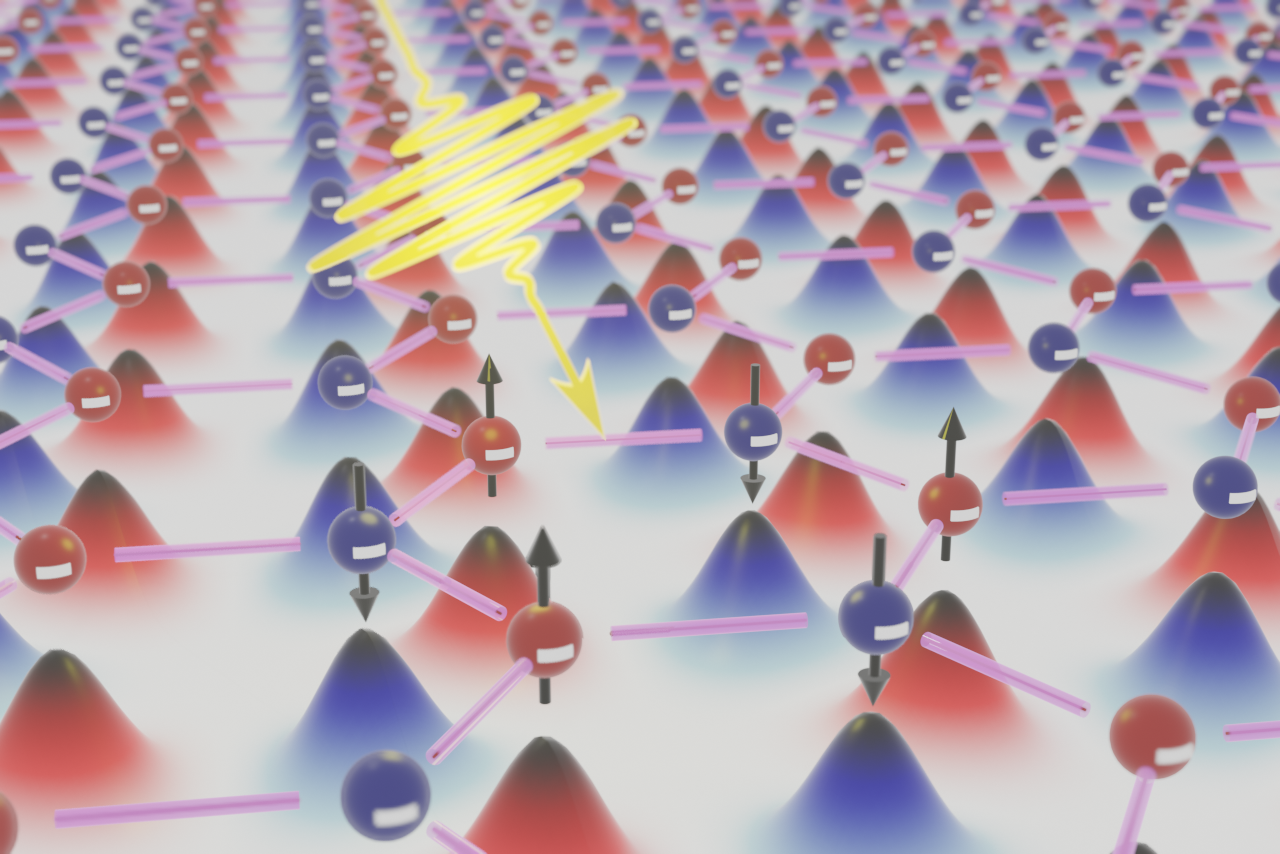}
\caption{Sketch of the 2D Wigner crystal with a honeycomb lattice and alternating spin polarization. The colored curves underneath the particles illustrate their wave functions. A low energy light field can probe the internal transition of the Wigner crystal towards excited quantum states to provide a conclusive proof of the phase transition. }
\label{fig:scheme} 
\end{figure}

In this work, we propose to exploit terahertz radiation to directly probe the internal quantum transitions of the WC in TMD monolayers, as illustrated in Fig. \ref{fig:scheme}. We have developed a fully quantum mechanical model of the interaction of monolayer WC with low frequency light. Our model is based on the many-particle density matrix formalism, where the Hamiltonian of interacting electrons is transformed into the self-consistent eigenbasis of the WC on a Hartree-Fock level. Consequently, we analyze the linear optical response of the system resulting from transitions between different bands of the electron crystal. 

We predict the emergence of a series of terahertz resonances whose position is a function of electron density and its characteristic peak sequence can be explained by the nature of the involved excited quantum states. Consequently, the predicted low frequency response can be used as a direct probe of the Wigner crystallization in arbitrary 2D systems of interacting fermions. We apply our microscopic model to determine the material realistic response of hBN-encapsulated WSe$_2$ monolayers without any open parameters and explain the characteristic response shape with the help of our microscopic model. We predict that the central energy of the internal quantum resonances shifts from about 4meV at $10^9$cm$^{-2}$ to 24meV at $8\cdot10^{10}$cm$^{-2}$. Moreover, we demonstrate that an enhanced effective mass and reduced dielectric screening, such as in MoSe$_2$ on SiO$_2$, can push the WC resonances up to 100meV. Finally, we also present a unitless study for ideal 2D systems, which can be rescaled to arbitrary fermionic systems.   


{\it{Microscopic mean field theory.}} - Previous numerical studies on WCs based on quantum Monte Carlo methods \cite{drummond2004diffusion} have demonstrated that the low density crystal phase is characterized by small correlation energies and that the WCs properties are well reproduced within a Hartree-Fock description of the Coulomb interaction \cite{trail2003unrestricted}. To obtain access to the optical response of the WC we start with the Hamilton operator of interacting electrons,
\begin{eqnarray}\label{eq:H_HF}
H=\sum_{\sigma \vec{k}} \varepsilon_{\vec{k}} a^\dagger_{\sigma \vec{k}} a_{\sigma \vec{k}} +\sum_{\sigma \vec{k q}} V^\text{HF}_{\sigma \vec{k}}(\vec{q})\, a^\dagger_{\sigma \vec{k-q}} a_{\sigma \vec{k}}, 
\end{eqnarray}
which can e.g. be related to the free charge carriers within n(p)-doped TMD monolayers, such that $a^{(\dagger)}_{\sigma \vec{k}}$ creates/annihilates electrons (holes) in the lowest conduction band (highest valence band) with spin up at the K point ($\sigma=\uparrow$) or spin down at the -K point ($\sigma=\downarrow$) and momentum $\vec{k}$. For low temperature and densities, the electron dispersion $\varepsilon_{\vec{k}}$ is determined by an effective mass $m_\ast$ \cite{kormanyos2015k}. The Coulomb interaction is treated in a mean field approximation\cite{trail2003unrestricted, pan2020quantum} via the non-local Hartree-Fock potential,
\begin{eqnarray}\label{eq:V_HF}
V^\text{HF}_{\sigma \vec{k}}(\vec{q})=\sum_{\sigma' \vec{k'}} ( V_\vec{q} -V_\vec{k-k'-q}\delta_{\sigma\sigma'})\, w^{\sigma'}_\vec{k'}(\vec{q}),
\end{eqnarray}
depending on the Fourier transform of the Coulomb potential $V_\vec{q}$, which includes the dielectric environment of the TMD monolayer through the Rytova-Keldysh potential \cite{rytova1967, brem2020phonon}. Moreover, the mean field potential also depends on the charge distribution entering via the Fourier transform of the Wigner distribution $w^{\sigma}_\vec{k}(\vec{q})=\langle a^\dagger_{\sigma \vec{k+q}} a_{\sigma \vec{k}} \rangle$. The Hamiltonian in Eq. (\ref{eq:H_HF}) is now diagonalized via the basis transformation $A^\dagger_{\lambda\sigma\vec{k}}=\sum_\vec{G}u^{\lambda\ast}_{\sigma\vec{k}}(\vec{G}) a^\dagger_{\sigma,\vec{k+G}}$, where we sum over the reciprocal lattice vectors $\vec{G}$ of the WC. Here,  $u^{\lambda}_{\sigma\vec{k}}(\vec{G})$ is the complete set of Bloch functions fulfilling the periodic Hartree-Fock equation (cf. SI). The latter is solved iteratively until a self-consistent solution is found,  starting with a randomized charge distribution. Hence, in the new basis describing WC electrons the Hamiltonian reads,
\begin{eqnarray}\label{eq:H_WC}
H=\sum_{\lambda\sigma\vec{k}} E^\lambda_{\sigma\vec{k}}A^\dagger_{\lambda\sigma\vec{k}} A_{\lambda\sigma\vec{k}},
\end{eqnarray}
where the self-consistent solution requires that the lowest band of the WC $\lambda=0$ is completely filled (one electron per WC unit cell) and the rest is empty, i.e. $f_{\lambda\sigma\vec{k}}=\langle A^\dagger_{\lambda\sigma\vec{k}} A_{\lambda\sigma\vec{k}}\rangle=\delta_{\lambda 0}$. 

Having determined the energy spectrum of WC electrons $E^\lambda_{\sigma\vec{k}}$ and their wave functions $u^{\lambda}_{\sigma\vec{k}}(\vec{G})$, we can transform the electron-light interaction into the Wigner basis and compute the dynamics of the WC using the Heisenberg's equation of motion\cite{lindberg1988effective,kira2006many}. In particular, assuming a weak perturbation through the electromagnetic field we can derive the linear response of the system, which e.g. can be specified via the diagonal components of the linear optical conductivity tensor reading
\begin{eqnarray}\label{eq:sigma}
\sigma(\omega)=\dfrac{i\hbar}{Ad} \sum_{\sigma\lambda\vec{k},\pm} \dfrac{\lvert J^{\lambda}_{\sigma \vec{k}}\rvert^2}{\Delta E^{\lambda}_{\sigma \vec{k}}}(\hbar\omega \pm \Delta E^{\lambda}_{\sigma \vec{k}} + i\Gamma)^{-1},
\end{eqnarray}
where we have introduced the monolayer thickness $d$ \cite{laturia2018dielectric}, the normalization area $A$ and a phenomenological dephasing constant $\Gamma$. The oscillator strength of the interband transition with energy $\Delta E^{\lambda}_{\sigma \vec{k}}=E^{\lambda}_{\sigma \vec{k}}-E^{0}_{\sigma \vec{k}}$ for $\hat{\vec{e}}$-polarized light is given by the current matrix element $J^{\lambda}_{\sigma \vec{k}}=e_0\hbar/m_\ast \sum_\vec{G} u^{\lambda\ast}_{\sigma\vec{k}}(\vec{G}) \vec{G}\cdot \hat{\vec{e}} \;u^{0}_{\sigma\vec{k}}(\vec{G})$. All details of the derivation and the used parameters are given in the SI. The microscopic model described above is not specific to TMD monolayers and can be applied to arbitrary 2D electron system. In principle, a similar approach can be used to compute the optical response also in generalized WCs within twisted bilayers.


\begin{figure}[t!]
\includegraphics[width=\columnwidth]{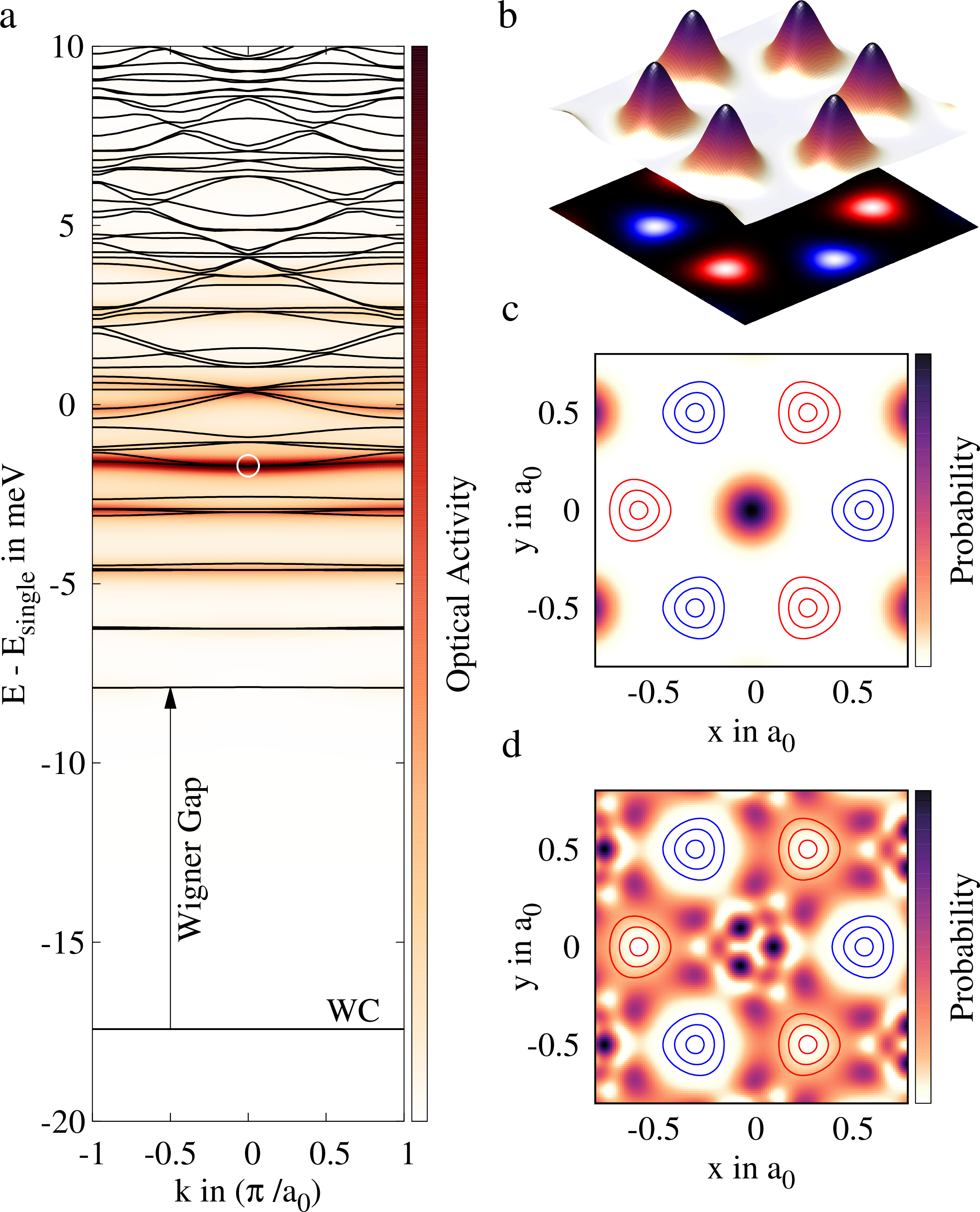}
\caption{Internal quantum structure of the WC in  p-doped, hBN-encapsulated WSe$_2$ at $r_s=50$. a) Band structure of the WC, where only the lowest band is fully occupied. The background color illustrates the oscillator strength for optical transitions from the ground state. b) The surface plot shows the charge density, whereas the projection contains the spin polarization ($\uparrow$ is red, $\downarrow$ is blue). The wave function  at $\vec{k}=0$ of the c) first excited state and d) the state with the largest oscillator.}
\label{fig:bands} 
\end{figure}
{\it{Internal quantum structure of WCs.}} - Now, we apply the developed model to the exemplary system of p-doped, hBN-encapsulated WSe$_2$ monolayers at a density of about $n\approx3\cdot 10^{10} \text{cm}^{-2}$ ($r_s=50$), i.e. deep within the crystalline phase that was predicted to melt at $r_s=30-40$ \cite{yoon1999wigner, drummond2009phase,zarenia2017wigner}. The temperature is assumed low enough such that only the lowest band of the WC is occupied and the thermal occupation of the excited states can be neglected.
Figure \ref{fig:bands} shows the computed band structure for electrons with $\sigma=\uparrow$ (on a cut through the WCs Brillouin zone) as well as the wave functions of the monolayer WC. The band structure is illustrated in Fig.\ref{fig:bands}a, where the background has been colored encoding the optical activity $I_\sigma(\vec{k},E)=\sum_{\lambda}\lvert J^{\lambda}_{\sigma \vec{k}}\rvert^2 \delta(E-E^{\lambda}_{\sigma \vec{k}})$. We find a completely flat ground state at about 17meV below the edge of the single particle dispersion. We also observe a series of flat excited states, where the first is separated from the ground state by a Wigner gap of about 10meV. With the increasing band index the excited states become more dense and dispersed. In order to interpret this result we investigate the underlying nature of these quantum states and their respective wave functions. 

The fully occupied ground state is in fact the WC itself, i.e. the corresponding wave functions are eigenstates of the Coulomb potential created by their own collective charge distribution. Here, the charge density (Fig.\ref{fig:bands}b) exhibits strongly localized peaks ordered in a honeycomb lattice \cite{bonsall1977some} with alternating spin polarization (red and blue projections). Moreover, the localized charge distributions exhibit a significant triangular warping which is usually neglected in variational approaches. 

In contrast to the self-consistent ground state (solution to the non-linear Hartree-Fock equation), the excited states are unoccupied and therefore simply the solutions of the Schrödinger equation with a non-local Coulomb potential determined by the ground state charge density. Figure \ref{fig:bands}c shows the wave function of the first excited state at $\vec{k}=0$, whereas the contours indicate the peaks of the $\uparrow$- (red) and $\downarrow-$ charge density (blue) from Fig.\ref{fig:bands}b. In order to minimize the repulsive Coulomb energy, the excited state is strongly localized in the pockets between the charge peaks. Consequently, the wave function overlap with the ground state is small, which is reflected by the weak optical activity of the first excited state. 
With the increasing band index, we first find several other localized states with increasing kinetic energy and growing spatial extent\cite{brem2020tunable}. Consequently, the localized orbitals of higher order bands begin to overlap at higher quantum indices and we find dispersed bands at high energies corresponding to scattering states. Figure \ref{fig:bands}d shows the excited state with the largest optical activity for transitions from the occupied ground state. As result of its delocalized character and the beneficial exchange interaction between equally spin-polarized electrons, the overlap with the ground state (red contours) is large, resulting in a strong current matrix element. 

The proposed detection mechanism via low-energy excitations has previously been applied to WCs formed in GaAs quantum wells in strong magnetic fields, where a single peak has been observed at microwave energies\cite{li2000microwave, chen2003microwave}, the so called pinning frequency\cite{chitra2001pinned}. In contrast, our microscopic model predicts that zero field monolayer Wigner crystals should exhibit a whole series of resonances at much larger frequencies.

With the knowledge of the internal WC quantum structure, we can now consider the dynamics of the system under external perturbations, e.g. the absorption of light.
\begin{figure}[t!]
\includegraphics[width=\columnwidth]{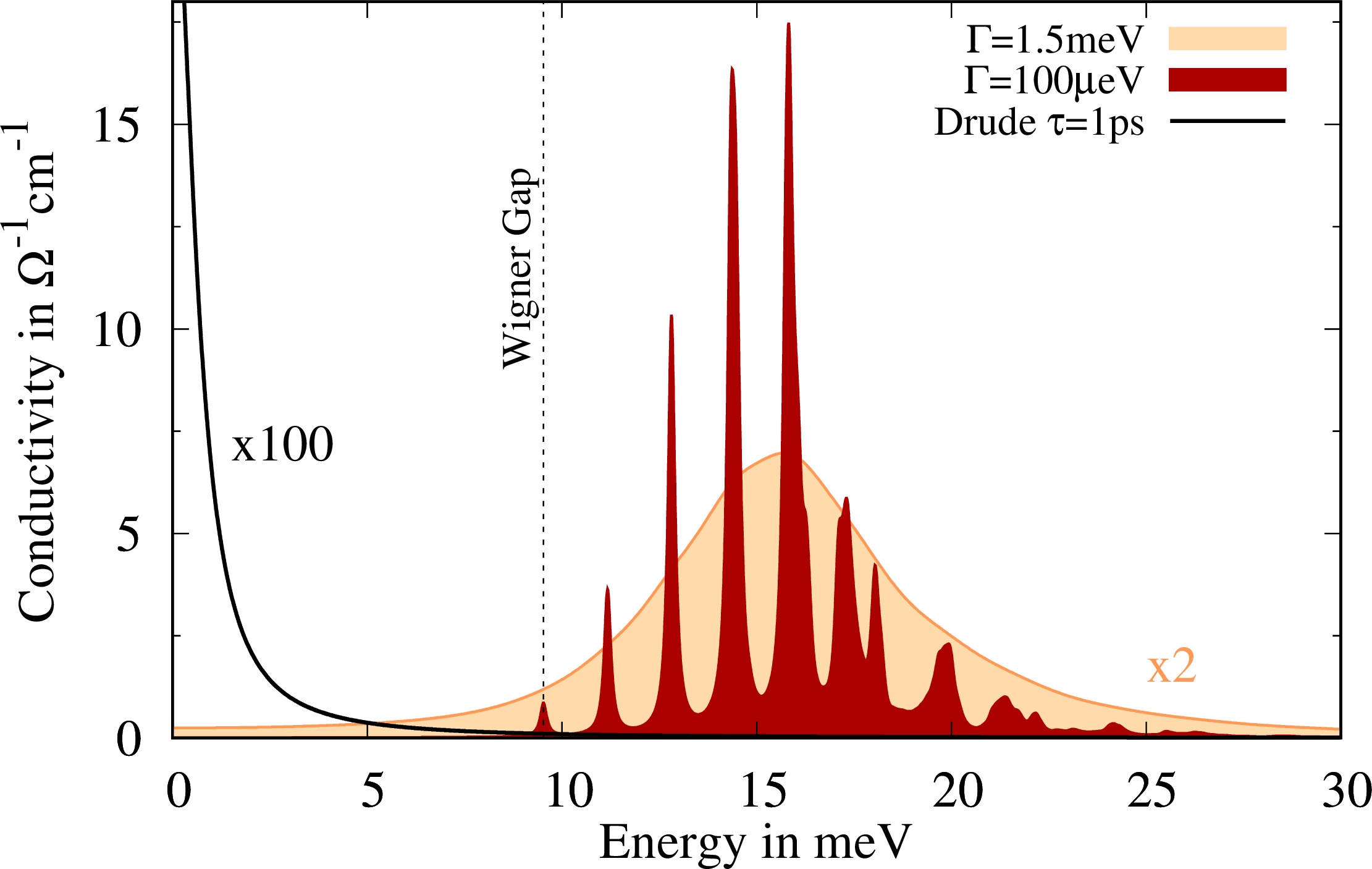}
\caption{Terahertz fingerprint of the Wigner crystal. The real part of the optical conductivity exhibits a characteristic series of internal quantum transitions, whose amplitude sequence can be explained by the nature of the involved excited states. Note that the strongly broadened response signal has been scaled by a factor of 2, whereas the Drude signal of free electrons (black line) has been multiplied by 100.}
\label{fig:spec} 
\end{figure}

{\it{Optical fingerprint of WCs.}} Now, we evaluate the linear optical conductivity from Eq. (\ref{eq:sigma}) in the energy range of the WCs interband transition. Figure \ref{fig:spec} illustrates the real part of the resulting spectrum for two exemplary linewidths.
For the narrower linewidth of $\Gamma=0.1$meV the spectrum exhibits a distinguished multi-peak structure with a characteristic sequence of peak amplitudes. The series of resonances starts at the energy of the Wigner gap (cf. Fig. 2a), reflecting the transition to the first excited state, which has a weak oscillator strength as discussed above. The subsequent series of peaks corresponds to the transition to higher order excited states, whose oscillator strength increases until a maximum is reached. This is a result of the growing kinetic energy of the excited states with increasing band index. With the increasing orbital size of the excited states, which are mostly localized in the pockets between the WC electrons, their overlap with the ground state electrons becomes enhanced. After a maximum is reached for the excited state illustrated in Fig. \ref{fig:bands}d, the oscillator strength decreases again. Here, the final states momentum spectrum $u^{\lambda}_{\sigma\vec{k}}(\vec{G})$ begins to move towards large $\vec{G}$ vectors, corresponding to scattering states with a large kinetic energy. Consequently, the momentum space overlap determining $J^{\lambda}_{\sigma \vec{k}}$ shrinks. 

For the larger linewidth of $\Gamma=1.5$meV the spectrum melts to a single peak centered at about 15meV. For comparison, we have also included the Drude response that is expected for an electron gas at the same density. This shows that even for broad resonances the terahertz spectrum is undergoing a dramatic change during the formation of the WC. However, we expect a much smaller linewidth then $\Gamma=1.5$meV, since the crystallization is taking place at very low temperatures and the phase-space for electron-phonon scattering is strongly suppressed within the flat bands of the WC. Therefore, we predict a clear multi-peak shape allowing to distinguish different transitions between the internal quantum states of the WC.

\begin{figure}[t!]
\includegraphics[width=\columnwidth]{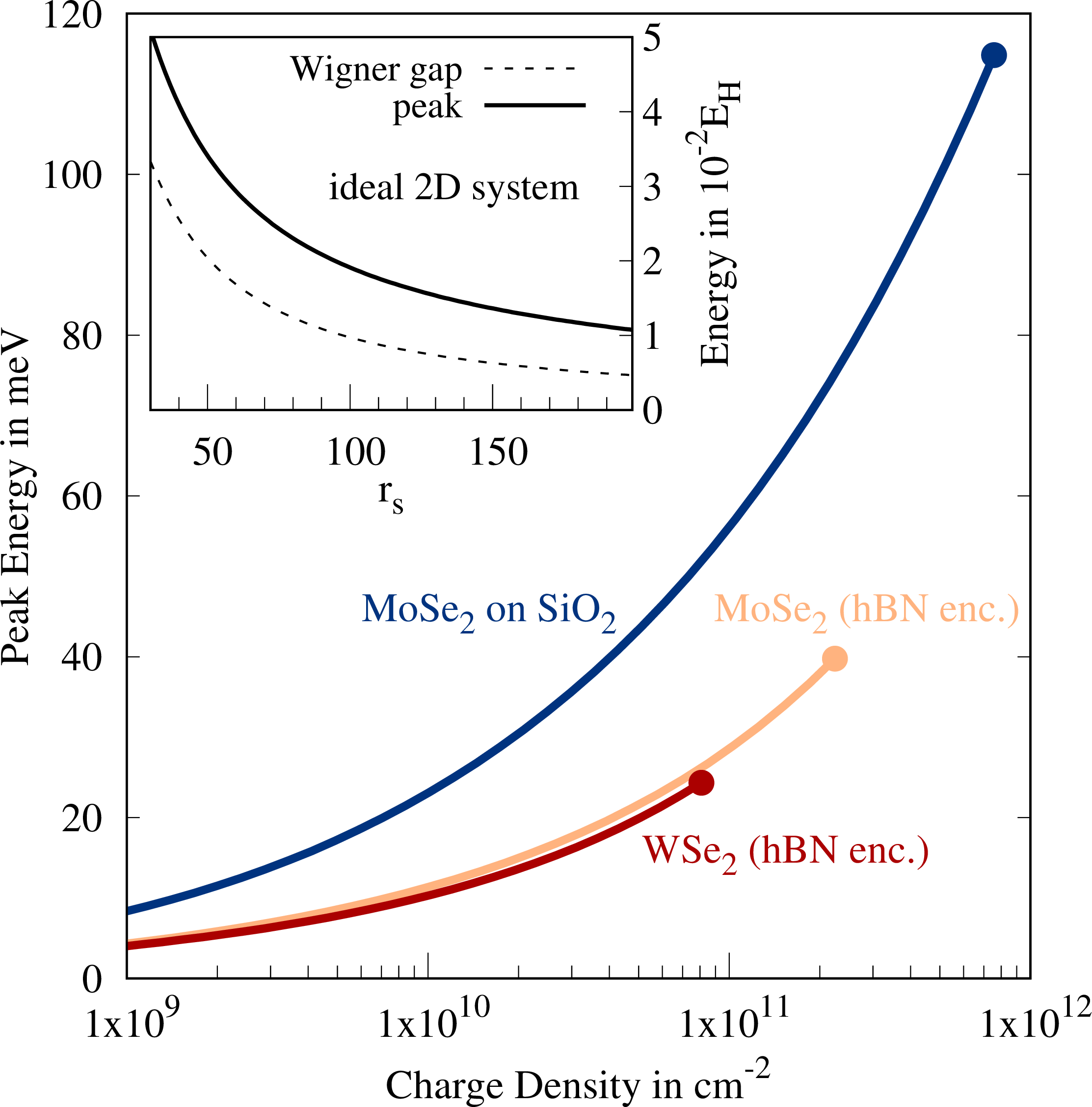}
\caption{Central energy of the optical response signal as function of the hole density for different monolayers/substrates. The inset shows the result for an ideal 2D system, where the only system parameter is given by $r_s$ and the energy scale is given by effective Hartree.}
\label{fig:scan} 
\end{figure}

{\it{Density dependence.}} Now,  we vary the charge density and investigate how the spectral response shifts as the lattice parameter of the WC $a_0=\sqrt{4/(\sqrt{3} n)}$ is changed. To this end we consider the center of the spectral response defined as $\omega_\text{peak}=\int  \omega\sigma(\omega)d\omega/\int \sigma(\omega)d\omega$, where the integrals are only performed over positive frequencies. Figure \ref{fig:scan} shows the center energy of the optical response as function of density for different TMD monolayers.
The filled dots represent the values at $r_s=30$, such that for higher densities quantum melting of the WC is expected. For all three illustrated systems we consider p-doping, since the valence band in TMDs has a significantly larger mass then the conduction band, which enables larger critical densities before quantum melting sets in. Figure \ref{fig:scan} illustrates that the optical response shifts significantly to higher energy at enhanced charge densities, which can serve as a clear indicator that the observed peaks indeed stem from transitions of the WC. With increasing density, the charge density peaks are more closely spaced which enhances the repulsive Coulomb energy for the excited states and therefore increases the transition energies. As a result, we also find that the transition energies for samples on SiO$_2$ are significantly larger, since the Coulomb interaction is much less screened by the substrate. In addition to the spectral shift of the center energy, the amplitude of the optical response also increases with growing densities, since the density of optically active particles is enhanced. 

We have additionally performed calculations for a ideal 2D system with $V_q\propto 1/q$, which reproduces the results for TMD monolayers at small densities. For an ideal 2D system the only relevant system parameter is given by $r_s=1/\sqrt{\pi n a_B^2}$ with the effective Bohr radius $a_B=4\pi \epsilon_0\epsilon \hbar^2/(m_\ast e_0^2)$. Using this density scale, the energies can be given in terms of effective Hartree $E_H=\hbar^2/(m_\ast a_B^2)$, such that the results given in the inset of Fig. \ref{fig:scan} can be rescaled to arbitrary 2D system. The inset furthermore demonstrates that the overall width of the optical response $\Delta E \approx E_\text{peak}-E_\text{gap}$ also grows with increasing density, since the distance between excited states becomes larger.

We want to emphasize that the applied model is only valid for weak excitation conditions, i.e. assuming a negligible population of the excited states. A lower bound for the critical power density of the exciting laser can be obtained by estimating a lower bound for the life time of excited states. Assuming that the life time is shorter than $\tau=1$ns, the weak excitation regime can be specified by a power density of $I \ll n (E_1-E_0)/\tau$, which for the system studied in Fig. \ref{fig:bands} and \ref{fig:spec} yields $I \ll 50$mW/cm$^2$. For stronger excitation powers we expect optical non-linearities to occur that could even result in a destruction of the WC.

In conclusion, our study demonstrates that the terahertz response of monolayer Wigner crystals can be used as an unambiguous fingerprint, exhibiting a characteristic sequence of internal quantum transitions. The inherent spectral shift of resonance frequencies with the charge density further represents a strong benchmark for the crystallization. We have predicted the peak positions for different TMD monolayers as well as other 2D fermion systems. The presented results will guide future experiments towards the detection of Wigner crystallization and the developed approach can be further exploited to theoretically study the interaction dynamics in pure as well as generalized Wigner crystals in twisted bilayers.

\begin{acknowledgments}
We acknowledge support from Deutsche Forschungsgemeinschaft (DFG) via SFB 1083 (Project B9) and the European Unions Horizon 2020 research and innovation program under grant agreement No 881603 (Graphene Flagship).
\end{acknowledgments}

\end{document}